\documentclass[twocolumn,showpacs,preprintnumbers,amsmath,amssymb,prb,floatfix]{revtex4}

\usepackage{graphicx}
\usepackage{bm}

\begin{document}

\title{Doping dependence of the superconducting gap in Tl$_2$Ba$_2$CuO$_{6+\delta}$ from heat transport}

\author{D. G. Hawthorn}
\altaffiliation[Present address: ]{Department of Physics and Astronomy, University of British Columbia, Vancouver, British Columbia, Canada.}
\affiliation{Department of Physics, University of Toronto, Toronto, Canada, M5S 1A7}

\author{S. Y. Li}
\altaffiliation[Present address: ]{D{\'e}partement de physique and RQMP, Universit{\'e} de Sherbrooke, Sherbrooke, Canada.}
\affiliation{Department of Physics, University of Toronto, Toronto, Canada, M5S 1A7}

\author{M. Sutherland}
\altaffiliation[Present address: ]{Cavendish Laboratory, University of Cambridge, Cambridge, U.K.}
\affiliation{Department of Physics, University of Toronto, Toronto, Canada, M5S 1A7}

\author{Etienne Boaknin}
\affiliation{Department of Physics, University of Toronto, Toronto, Canada, M5S 1A7}

\author{R. W. Hill}
\altaffiliation[Present address: ]{Department of Physics, University of Waterloo, Waterloo, Canada.}
\affiliation{Department of Physics, University of Toronto, Toronto, Canada, M5S 1A7}

\author{C.~Proust}
\altaffiliation[Present address: ]{Laboratoire National des Champs Magn{\'e}tiques Puls{\'e}s, Toulouse, France.}
\affiliation{Department of Physics, University of Toronto, Toronto, Canada, M5S 1A7}

\author{F. Ronning}
\altaffiliation[Present address: ]{Los Alamos National Lab, Los Alamos, USA.}
\affiliation{Department of Physics, University of Toronto, Toronto, Canada, M5S 1A7}

\author{M. A. Tanatar}
\altaffiliation[Present address: ]{D{\'e}partement de physique and RQMP, Universit{\'e} de Sherbrooke, Sherbrooke, Canada}
\altaffiliation[Permanent address: ]{Inst. Surf. Chem., N.A.S. Ukraine.}
\affiliation{Department of Physics, University of Toronto, Toronto, Canada, M5S 1A7}

\author{Johnpierre Paglione}
\altaffiliation[Present address: ]{Department of Physics, University of California, San Diego, La Jolla, USA.}
\affiliation{Department of Physics, University of Toronto, Toronto, Canada, M5S 1A7}

\author{Louis Taillefer}
\email{Louis.Taillefer@USherbrooke.ca} 
\altaffiliation[Present address: ]{D{\'e}partement de physique and RQMP, Universit{\'e} de Sherbrooke, Sherbrooke, Canada}
\affiliation{Department of Physics, University of Toronto, Toronto, Canada, M5S 1A7} 

\author{D. Peets}
\affiliation{Department of Physics and Astronomy, University of British Columbia, Vancouver, Canada, V6T 1Z4}
 
\author{Ruixing Liang}
\affiliation{Department of Physics and Astronomy, University of British Columbia, Vancouver, Canada, V6T 1Z4}

\author{D. A. Bonn}
\affiliation{Department of Physics and Astronomy, University of British Columbia, Vancouver, Canada, V6T 1Z4}
 
\author{W. N. Hardy}
\affiliation{Department of Physics and Astronomy, University of British Columbia, Vancouver, Canada, V6T 1Z4}
 
\author{N. N. Kolesnikov}
\affiliation{Institute of Solid State Physics, Russian Academy of Sciences, Chernogolovka, Russia}

\date{\today}

\begin{abstract}
We present low-temperature thermal conductivity measurements on the cuprate $\rm{Tl_2Ba_2CuO_{6+\delta}}$ throughout the overdoped regime.  In the $T \rightarrow 0$ limit, the thermal conductivity due to $d$-wave nodal quasiparticles provides a bulk measurement of the superconducting gap, $\Delta$.  We find $\Delta$ to decrease with increasing doping, with a magnitude consistent with spectroscopic measurements (photoemission and tunneling).  
This argues for a pure and simple $d$-wave superconducting state in the overdoped region of the phase diagram, 
which appears to extend into the underdoped regime down to a hole concentration of $\simeq 0.1$~hole/Cu.  
As hole concentration is decreased, the gap-to-$T_c$ ratio increases, showing that the suppression of the superconducting transition temperature $T_c$ (relative to the gap) begins in the overdoped regime.
\end{abstract}

\pacs{74.25.Fy,74.72.Jt}

\maketitle

Over the past decade, low-temperature thermal conductivity has emerged as a powerful probe of the low-energy quasiparticle excitations in the cuprates.  Unlike conventional $s$-wave superconductors, the nodes in the $d$-wave superconducting order parameter of cuprates lead to a finite quasiparticle density of states at zero energy resulting in ``universal'' transport in the low-temperature limit ($T \rightarrow 0$).\cite{Taillefer97,Nakamae01}  By performing measurements in the $T \rightarrow 0$ limit, the complicated physics of quasiparticle transport in these materials is greatly simplified and meaningful quantitative comparisons between mean-field theories in the self-consistent T-matrix approximation (SCTMA) \cite{Lee93,Graf96,Durst00} and experiments have been made.\cite{Chiao00,Proust02,Sutherland03}  

In the mean-field theory, the low-$T$ thermal conductivity is proportional to the slope of the gap near the nodes, $\delta \Delta /\delta \phi |_{\phi \rightarrow \pi/4}$.  By assuming a gap with a simple $d$-wave form ($\Delta=\Delta_0\cos(2\phi)$), the gap maximum, $\Delta_0$, can be inferred from $\delta \Delta /\delta \phi$, making thermal conductivity a measure of the superconducting gap.  This measure is notable for a number of reasons that make it an excellent complement to more conventional, spectroscopic measurements of the gap.  First, it is a very low energy measurement of the gap, the energy scale being set by temperature (and impurity bandwidth, $\gamma$).  Second, it measures the gap near the nodes (in the $\Gamma \rightarrow (\pi,\pi$) direction) as opposed to the anti-nodes ($\pi$,0).   Finally, thermal conductivity provides a bulk measure of the gap, unlike surface sensitive techniques such as photoemission (ARPES) and tunneling.  Together these properties allow one to impose important constraints on the nature of superconductivity in cuprates.

The cuprate Tl$_2$Ba$_2$CuO$_{6+\delta}$ (Tl-2201) is an ideal material for quantitative investigations of $d$-wave superconductivity in the cuprates.  It is a single-layer material with $T_c^{max} \simeq 90$ K, comparable to the 
well-studied cuprates Bi$_2$Ba$_2$CaCu$_2$O$_{8+\delta}$ (Bi2212) and YBa$_2$Cu$_3$O$_y$ (YBCO).  By varying the oxygen concentration, the material can be doped from optimal doping far into the overdoped region of the phase diagram.  
In this region, the metallic state is well-characterised in terms of a single coherent Fermi surface and there are no indications of competing phases, such as encountered in La$_{2-x}$Sr$_x$CuO$_4$ (LSCO), for example.
Moreover, Tl-2201 can be prepared with reasonably low levels of disorder because the dopant atoms are interstitial oxygen residing well away from the CuO$_2$ planes, instead of the more obtrusive cation doping located close to the CuO$_2$ planes (as in LSCO, for example). 

\begin{figure}[th]
\centering
\resizebox{\columnwidth}{!}{\includegraphics{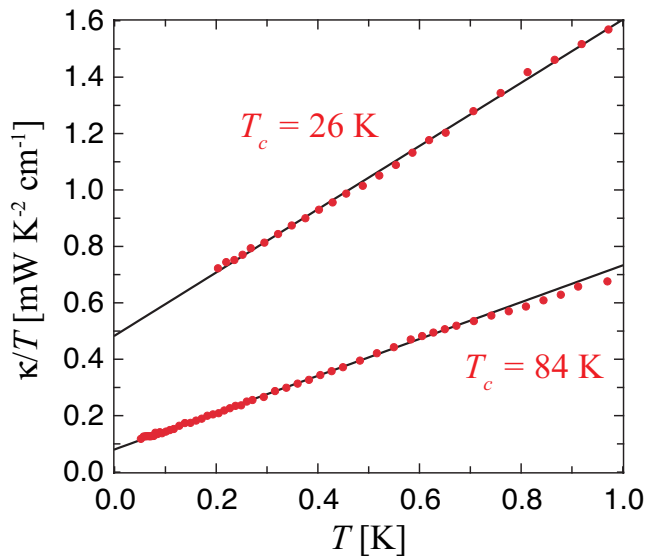}}
\caption[$\kappa/T $ vs. $T$ below 1 K for Tl-2201 samples at various doping levels.]{\label{fig1}(colour online) Thermal conductivity of one Tl-2201 sample, measured at two different hole concentrations, as indicated by the respective $T_c$ values, plotted as $\kappa/T $ vs $T$ below 1 K. The lines are fits to Eq.~\ref{eq:kappafitTl}.
A three-fold decrease in $T_c$ corresponds to a six-fold increase in the residual linear term (which is inversely proportional to the superconducting gap, via Eq.~\ref{eq:koTvsDelta}),
hence a two-fold decrease in the gap-to-$T_c$ ratio with doping.}
\end{figure}

\begin{figure}[th]
\centering
\resizebox{\columnwidth}{!}{\includegraphics{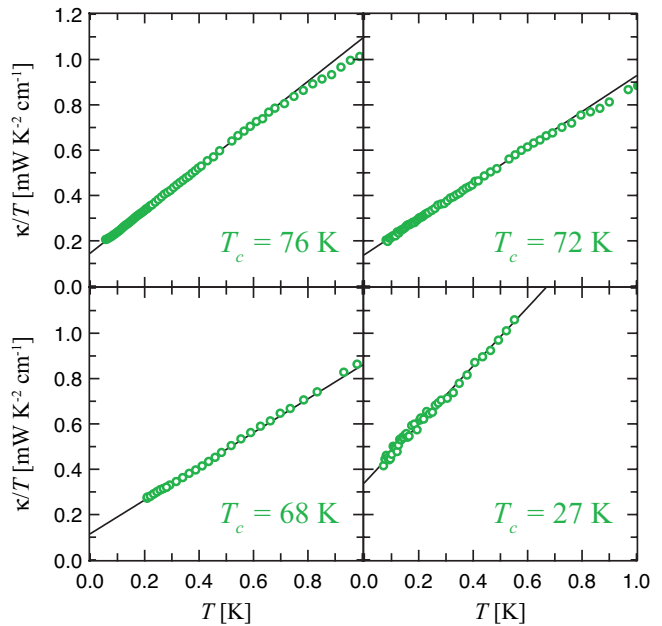}}
\caption[$\kappa/T $ vs. $T$ below 1 K for Tl-2201 samples at various doping levels.]{\label{fig2}(colour online) Thermal conductivity of four Tl-2201 samples, each with a different hole concentration, as indicated by their respective $T_c$ values, plotted as $\kappa/T $ vs $T$ below 1 K. The lines are fits to Eq.~\ref{eq:kappafitTl}.}
\end{figure}

In this paper, we present low-temperature thermal conductivity measurements of Tl-2201 as a function of doping, spanning the overdoped region of the phase diagram.  Expanding upon previous work on cuprates,\cite{Taillefer97,Chiao00,Nakamae01,Proust02,Sutherland03} we analyze our results using the mean-field SCTMA theory to provide a determination of the superconducting gap from thermal conductivity.  We find agreement between thermal conductivity and spectroscopic measurements of the gap, confirming both the $d$-wave form of the superconducting gap as well as the validity of the mean-field description of low-temperature thermal conductivity in overdoped Tl2201.  We track the gap as a function of doping throughout the overdoped region and find a gap-to-$T_c$ ratio, $\mu \equiv \Delta_0/k_BT_c$, that increases with decreasing doping, starting with a value close to the weak-coupling BCS value ($\mu$ = 2.14) at high doping and reaching a value twice as large near optimal doping ($\mu = 5 - 6$).  This shows that the suppression of the onset of coherent superconductivity (or $T_c$) so characteristic of the underdoped regime actually begins above optimal doping.

Five flux-grown single crystals of $\rm{Tl_2Ba_2CuO_{6+\delta}}$ are used in this study.\cite{Peets07,Kolesnikov95}  The platelet samples have approximate length and width of $\sim 0.5$ mm in the $ab$ plane and thickness $\times$ 20 - 65 $\mu$m along the $c$ axis.  The heat current is applied in the $ab$ plane and the magnetic field along the $c$-axis. The concentration of holes in the CuO$_2$ plane, $p$, is set by tuning the excess oxygen content in the TlO$_2$ layer.  
Concentrations between optimal doping and heavily overdoped were achieved by annealing the samples in atmospheres ranging from vacuum ($\sim 5 \times 10^{-7}$ torr) at 500 $^\circ$C near optimal doping to 50-60 mbar oxygen at 400-480 $^\circ$C for heavily overdoped samples.  
The value of $p$ for each sample was estimated from the measured $T_c$ (defined as the temperature at which the electrical resistivity has fallen to zero), using the well-known empirical formula\cite{Presland91}
\begin{equation}
\frac{T_c}{T_c^{max}}=1-82.6(p-0.16)^2,
\label{eq:Tallon}
\end{equation}
where $T_c^{max}$ = 90 K is taken as the nominal $T_c$ of optimally-doped Tl-2201.
The value of $T_c$ and $p$ for each sample is listed in Table~I.

Thermal conductivity, $\kappa$, was measured down to 50 mK in a dilution refrigerator using a standard 4-wire steady-state technique described elsewhere.\cite{Sutherland03}
\renewcommand\thefootnote{\alph{footnote}}
Electrical contacts to the samples were made using evaporated gold pads, which were annealed to improve the contact conductance in the same anneal used to set the oxygen content.  Silver wires were then attached to the contact pads using silver paint.
\begin{table}
\begin{minipage}[t]{\columnwidth}
\begin{flushleft}
\caption{Residual linear term, $\kappa_0/T$, in the thermal conductivity of Tl-2201, from fits of $\kappa/T$ vs $T$ to Eq.~\ref{eq:kappafitTl}, for five single crystals with hole concentration $p$ estimated from measured $T_c$.  The gap maximum, $\Delta_0$, is evaluated from $\kappa_0/T$ using Eq.~\ref{eq:koTvsDelta}.  The error applies to $\kappa_0/T$, $\Delta_0$ and $\Delta_0/k_BT_c$.}
\label{table:TlkappaoT}
\begin{tabular}{|c|c|c|c|c|c|}
\hline 
~$T_c$~ & ~$p$~ & ~$\kappa_0/T$~ & ~$\Delta_0$~ & ~$\Delta_0/k_BT_c$~ & ~error~\\
~(K)~ &  & ~(mW K$^{-2}$ cm$^{-1}$)~ & ~(meV)~ & & ~(\%)~\\
\hline
~84\footnotemark[1]~ & ~0.188~ & 0.08 & 40 & 5.6 & 17\\
76~ & 0.203 & 0.15 & 22 & 3.4 &16\\
72~ & 0.209 & 0.14 & 23 & 3.8 & 15\\
68~ & 0.214 & 0.12  &  28 & 4.8 & 25\\
27~ & 0.252 & 0.34  & 9.5 & 4.1 & 24\\
26\footnotemark[1]& 0.253 & 0.48 & 6.7 & 3.0 & 17\\
\hline
\end{tabular}
\footnotetext[1]{Denotes the same sample measured at two dopings.}
\end{flushleft}
\end{minipage}
\end{table}
\renewcommand\thefootnote{\arabic{footnote}}

The largest error in the magnitude of $\kappa$, ranging from 15 to 25 \% (see Table~I), comes from the measurement of sample dimensions for the appropriate geometric factor. In order to eliminate this dominant source of uncertainty when comparing two different hole concentrations, one of the samples was measured twice with the same contacts, at two different dopings, $T_c$ = 84 K and 26 K (see data in Fig.~1).  This was achieved by first evaporating gold pads onto the sample and annealing the sample to set the doping, as described above.  After measuring the sample, the wires and silver paint were removed, the sample was reannealed to a different doping and subsequently re-measured with the gold pads still in place. 

The thermal conductivity measured below 1~K is shown in Fig.~\ref{fig1} for this sample and in Fig.~\ref{fig2} for four other samples, each at a different doping.   
To extract the electronic contribution to $\kappa$ the data are fit to

\begin{equation}
\frac{\kappa}{T} = \frac{\kappa_0}{T} + AT,
\label{eq:kappafitTl}
\end{equation}
where $\kappa_0/T$ (intercept) is the electronic contribution to the total thermal conductivity and the second term (slope) is the phonon contribution, $\kappa_{ph}/T = A T$.  The values of $\kappa_0/T$ from fits to Eq.~\ref{eq:kappafitTl} are listed in Table~I, along with the error bars for each sample, estimated from the statistical error in the fits ($< 1.5\%$) and the error in the geometric factor ($15 - 25\%$).  These results can be compared to a previous study on a strongly overdoped Tl-2201 crystal,\cite{Proust02} with $T_c$ = 15 K, which gave $\kappa_0/T = 1.4$ mW K$^{-2}$ cm$^{-1}$ and an excellent fit to Eq.~\ref{eq:kappafitTl}.

Note that in these overdoped samples $\kappa_{ph} \propto T^2$ (over a decade in temperature).  This is in contrast to previous reports of the low-temperature thermal conductivity of cuprates, measured typically at or below optimal doping, where $\kappa_{ph}\propto T^{2.5-3}$, consistent with phonons being scattered from the boundaries of the sample.\cite{Taillefer97,Chiao00,Nakamae01,Takeya02,Sutherland03,Hawthorn03}  In overdoped Tl-2201, with its higher carrier concentration, we attribute the $T^2$ dependence of $\kappa_{ph}$ to dominant electron-phonon scattering, known to yield $\kappa_{ph} \propto T^2$ at low $T$ in a metal.\cite{Butler78} Smith has shown that this is indeed the theoretically expected dependence of phonon transport in a $d$-wave superconductor,\cite{Smith05} in the range 0.1 - 1.0~K.
We emphasize that the main results of this paper, which stem from the doping dependence of $\kappa_0/T$, are unaffected by alternative methods to fit $\kappa_{ph}$, such as a power-law with adjustable power ($\kappa/T = a + bT^{\gamma}$).  

\begin{figure}[t]
\centering
\resizebox{3.0in}{!}{\includegraphics{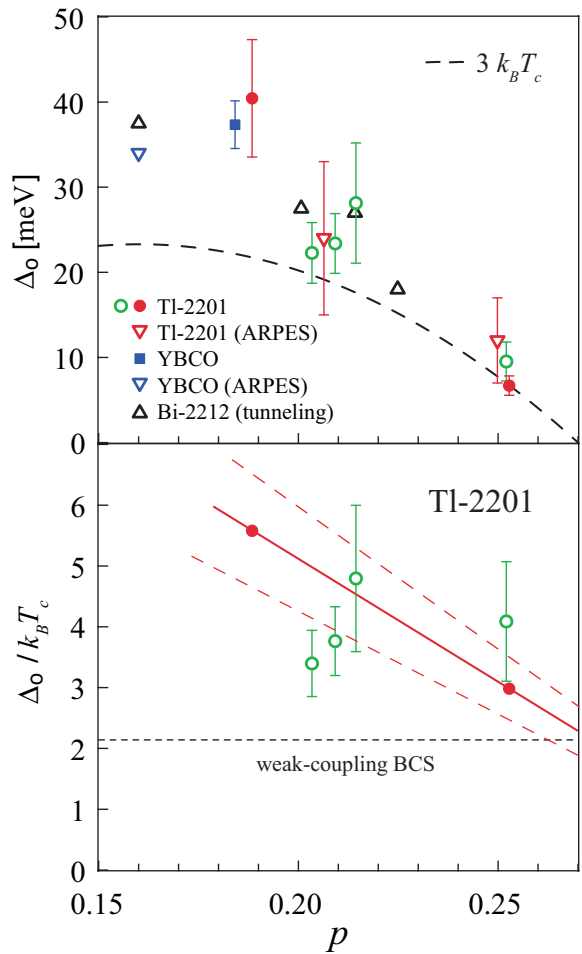}}
\caption[The gap maximum, $\Delta_0$, and gap ratio $\Delta_0/k_BT_c$ vs hole concentration, $p$, derived from thermal conductivity.]{\label{fig:DeltaTl}(colour online)  
{\it Upper panel}: The gap maximum, $\Delta_0$, vs hole concentration, $p$, derived from thermal conductivity measurements on Tl-2201 (circles) and YBCO (square; Refs.~\onlinecite{Sutherland03} and \onlinecite{Hill04}).   
The solid (red) circles denote the same Tl-2201 sample measured at two different doping levels, as discussed in the text.  Spectroscopic measurements of the gap from ARPES on Tl-2201 (red down-triangle; Refs.~\onlinecite{Plate05} and \onlinecite{Peets07}) and YBCO (blue down triangle; Ref.~\onlinecite{Nakayama06}) and SIS tunneling on Bi-2212 (black up-triangles; Ref.~\onlinecite{Miyakawa99}) are also shown.  The dashed line gives the BCS doping dependence of the gap based on the measured $T_c$, namely $\Delta_0 = \mu k_BT_c$, using a strong-coupling value of $\mu$ = 3 and $T_c(p)$ from Eq.~\ref{eq:Tallon}.
{\it Lower panel}:   The gap-to-$T_c$ ratio, $\Delta_0/k_BT_c$, vs $p$, for Tl-2201 from thermal conductivity data. 
The solid (red) line is a fit through the two data points (full red circles) obtained on a single sample, for which the relative uncertainty is minimal ($\sim 3$~\%), given the fixed geometric factor.  
The dashed (red) lines on either side reflect the uncertainty on the absolute value ($\pm 17$~\%; see Table~I). 
The horizontal (black) dashed line is the weak-coupling $d$-wave BCS prediction, namely $\Delta_0/k_BT_c$ = 2.14.\cite{Won94}}
 \end{figure}
 
The $T$-linear electronic contribution to $\kappa $ is due to transport from $d$-wave nodal quasiparticles in the  limit of $k_BT \ll \gamma \ll \Delta_0$, where $\gamma $ is the impurity band width.\cite{Graf96,Taillefer97,Durst00}  
In this limit, the residual linear term in the thermal conductivity is {\it universal}, in the sense that it is independent of scattering rate, and given by:\cite{Durst00} 

\begin{equation}
 \frac{\kappa_0}{T} = \frac{k_B^2}{3 \hbar} \frac{n}{c} \left( \frac{v_F}{v_{\Delta}} +
 \frac{v_\Delta}{v_F} \right),
 \label{eq:koT}
\end{equation}
where $n$ is the number of $\rm{CuO_2}$ planes per unit cell (=~2 for Tl-2201) and $c$ is the $c$-axis
lattice constant (=~23.2~\AA ~for Tl-2201).
The ratio $v_F / {v_{\Delta}}$ is the ratio of quasiparticle velocities normal and tangential to the Fermi surface at the node, respectively given by the Fermi velocity in the  (0,0) - $(\pi,\pi)$ direction, $v_F$, and

\begin{equation}
    v_{\Delta} =\frac{1}{\hbar k_F}\frac{d\Delta}{d\phi}\Big|_{node},
\label{eq:v2}
\end{equation}
proportional to the slope of the gap at the node.  Note that, unique amongst transport properties, $\kappa_0/T$ was shown to be robust against both Fermi liquid and vertex corrections.\cite{Durst00}  

The above formalism, however, is only valid in the limit $k_BT \ll \gamma \ll \Delta_0$.  Before proceeding, it is important to determine whether this criterion is indeed satisfied for our samples, particularly since $\Delta_0$ decreases with increasing doping and eventually goes to zero at the superconductor/metal phase transition.\cite{endnotegamma} 
In the Appendix, we provide two independent and consistent estimates of $\gamma$ for each sample which 
show that in all cases $\gamma / k_B T > 10$ (for $T < 1$~K) and $\gamma /\Delta_0 < 0.16$, validating the use of Eq.~\ref{eq:koT} for our Tl-2201 samples.

The significance of Eq.~\ref{eq:koT} is that it provides a reliable way to measure the gap from a rather straightforward measurement of the thermal conductivity.
Indeed, through Eqs.~\ref{eq:koT} and \ref{eq:v2}, the slope of the gap at the node, $\delta \Delta / \delta \phi |_{node}$, is immediately obtained from $\kappa_0 / T$, provided $v_F$ and $k_F$ are known.
\cite{Chiao00,Sutherland03}  
Making the assumption that $\Delta $ has the standard $d_{x^2-y^2}$ form $\Delta(\phi)=\Delta_0\cos(2\phi)$, Eq.~\ref{eq:koT} can be rewritten (for $v_F \gg {v_{\Delta}}$) as

\begin{equation}
\frac{\kappa_0}{T} \simeq \frac{k_B^2}{6} \frac{n}{c}k_F\frac{v_F}{\Delta_0}.
\label{eq:koTvsDelta}
\end{equation}
$\kappa_0/T$ can thus be related in a straightforward fashion to the gap maximum, $\Delta_0$, by making use of the fact that $v_F$ and $k_F$ at the nodes are approximately doping independent.  ARPES measurements\cite{Zhou03} have shown $v_F$ to be independent of both doping and material and to have a value $v_F$ = 2.7 $\times 10^7 \pm 20\%$ cm/s.   Measurements of the Fermi surface by ARPES in LSCO~\cite{Ino00}, Bi-2212~\cite{Ding97} and Tl-2201~\cite{Plate05}, show $k_F$ to have a value $\sim 0.7$~\AA$^{-1}$ ~along the nodal direction (0,0) - ($\pi$,$\pi$), essentially independent of material or doping ($k_F$ only has significant doping dependence along (0,0) - (0,$\pi$)).
In overdoped Tl-2201, this value of $k_F$ was nicely confirmed by bulk sensitive angular magneto-resistance oscillation (AMRO) measurements.\cite{Hussey03B}
The values of $\Delta_0$ obtained for our five Tl-2201 samples from Eq.~\ref{eq:koTvsDelta} are listed in Table~I, and plotted vs $p$ in the top panel of Fig.~\ref{fig:DeltaTl}.
Also shown is the value of $\Delta_0$ for an ultra-pure overdoped YBCO crystal,\cite{Sutherland03,Hill04} obtained in the same way (with $\kappa_0/T$ = 0.16 mW~K$^{-2}$ cm$^{-1}$) --
it is in excellent quantitative agreement with the Tl-2201 data.

$\Delta_0$ decreases with increasing doping and follows the doping dependence of the gap measured by spectroscopic probes such as tunneling~\cite{Renner98,Miyakawa99} and ARPES~\cite{White96}. In Figure~\ref{fig:DeltaTl} we also plot the doping dependence of $\Delta_0$ in Bi-2212 measured using SIS tunneling~\cite{Miyakawa99} and find good quantitative agreement with $\Delta_0$ measured by thermal conductivity,\cite{endnoteDelta} similar to previous work on optimally doped Bi-2212.\cite{Chiao00}  Comparing directly to measurements on Tl-2201, recent ARPES measurements give $\Delta_0 \simeq12 \pm 5$ meV and $\Delta_0 \simeq 24\pm 9$ meV in a samples with $T_c = 30$ K and 74 K respectively \cite{Plate05,Peets07} which is also in favorable agreement with thermal conductivity.  
($\Delta_0$ here is taken as the average of two commonly used measures of $\Delta_0$; the peak position and the leading-edge midpoint, which are 17 meV (33 meV) and 8 meV (15 meV), respectively, for $T_c = 30$ K (74 K).)

The agreement between the different measures of $\Delta$ has important implications.  
First, it provides further compelling evidence for coherent nodal quasiparticle excitations that are correctly described by mean-field $d$-wave quasiparticle theory throughout the overdoped region of the phase diagram.
Secondly, the fact that the gap derived from thermal conductivity (i.e. from low-energy {\it nodal} quasiparticles) agrees well with the gap maximum derived from tunneling or ARPES (determined from {\it anti-nodal} quasiparticles at higher energies) suggests that the gap structure (as a function of angle) is not very different from the simple $d$-wave gap assumed in our analysis ($\Delta_0$cos(2$\phi$)).  
This places limits on any sub-dominant order parameter that may exist in the bulk.  
%
Subdominant order parameters that result in a fully gapped nodal region  (such as $is$ or $id$) are limited to having a magnitude much less than the impurity bandwidth $\gamma$,\cite{Gusynin04} which is estimated to be on the order of $\Delta_0 / 10$ in our Tl-2201 samples (see Appendix).  An even more stringent restriction is provided by the YBCO measurements, where, in crystals of the highest quality, $\gamma$ can be orders of magnitude smaller than in Tl-2201.\cite{Hill04,Sutherland03} These conclusions are in agreement with the phase-sensitive tricrystal experiments, which show $d_{x^2-y^2}$ symmetry of the order parameter and no signature of a subdominant {\em imaginary} contribution to the superconducting order parameter away from optimal doping.\cite{Tsuei04}
(Note that a {\em real} $s$ component \cite{Kirtley06} is allowed.)

The gap-to-$T_c$ ratio, $\mu = \Delta_0/k_BT_c$, is plotted vs $p$ in Figure~\ref{fig:DeltaTl}.  
$\mu$ is seen to decrease with increasing doping from a strong-coupling BCS value of $\mu$ = 5-6 near optimal doping towards the weak-coupling value of $\mu = 2.14$ at high doping.  This trend is made compelling by comparing measurements on the same sample at two different dopings (solid red circles), for which the relative uncertainty is small. Indeed for sample $a$ (see Table I), $\mu$ drops by a factor 1.9 (from 5.6 to 3.0) when the doping is increased from $p=0.19$ ($T_c = 84$~K) to $p=0.25$ ($T_c = 26$~K). 

Having established how the superconducting gap grows with  decreasing doping in
the overdoped regime of the phase diagram, it is is natural to ask how it
continues to evolve below optimal doping, into the enigmatic underdoped
regime.
Heat transport measurements on YBCO have shown that $\kappa_0/T$ decreases
monotonically, at least down to $p \simeq 0.1$.\cite{Sutherland03,Sun04,Sutherland05}
Based on Eq.~5, this was interpreted in terms of a gap $\Delta$ that keeps
on rising as $p$ is reduced.\cite{Sutherland03}
Note, however, that recent measurements of the in-plane resistivity in the
(field-induced) normal state of YBCO reveal a ``metal-insulator''
crossover at $p \simeq 0.09$.\cite{Proust06}
One should not expect the standard BCS analysis (leading
to Eq.~3) to apply below that crossover.
(The same situation was encountered in LSCO,\cite{Sutherland03} setting
in at a considerably higher doping since the ``metal-insulator'' crossover
in that system is at $p  \simeq  0.17$.\cite{Boebinger96})
The crossover to a non-metallic normal state may account for the
qualitative change observed in the behaviour of $\kappa_0/T$ when $p$ is
reduced to 5-6\%, where it becomes independent of both doping and magnetic field.\cite{Doiron-Leyraud06}
This overall evolution of $\kappa_0/T$ vs $p$ in YBCO is 
corroborated by recent ARPES measurements on Bi-2212, which show the gap near the nodes to have comparable magnitude and to rise with decreasing $p$ until $p \simeq 0.08$, and then fall abruptly.\cite{Tanaka06}

In conclusion, we have measured the doping dependence of the low-temperature thermal conductivity in the overdoped cuprate Tl-2201.  Analyzing the magnitude of $\kappa_0/T$ using mean-field theory, we show quantitative agreement between the slope of the gap at the nodes from thermal conductivity and spectroscopic measurements of the gap maximum at the anti-nodes in overdoped cuprates.  These measurements provide further confirmation of the applicability of mean-field SCTMA transport theory in the cuprates, and more generally the basic description of overdoped cuprates as $d$-wave superconductors.  As a function of reduced doping, the gap-to-$T_c$ ratio increases by 
nearly a factor 2 from $p=0.25$ to $p=0.19$. This reveals that the strong 
suppression of phase coherence (or $T_c$) characteristic of the underdoped 
regime, presumably caused by a combination of pair-breaking and phase 
fluctuations, in fact begins in the overdoped region, at least as high as 
$p \simeq 0.2$, perhaps higher.


\appendix
\section{Estimates of the impurity bandwidth}

In this Appendix we estimate the impurity bandwidth, $\gamma$, using two approaches: from the normal state transport and from the magnetic field dependence of $\kappa_0/T$ in the superconducting state.  Both approaches yield similar values of $\gamma$ which in all samples satisfy the two criteria for the validity of a ``universal'' limit analysis, namely $\gamma \ll \Delta_0$ and $T \ll \gamma$.

\subsection{From normal state transport}

First, we estimate $\gamma$ from the normal state quasiparticle scattering rate, $\Gamma_N$.  
Because the normal state of our samples could not be reached with the magnetic field available in our lab (15~T), we turn to a previous study \cite{Proust02} on a crystal with a similar quality but significantly lower $T_c$ (higher doping), namely (15~K). In that crystal, $\Gamma_N$ was estimated from $\kappa_N/T$ to be 0.52 meV.\cite{Proust02}  Using this value as representative of $\Gamma_N$ in our samples, we can estimate an upper bound on $\gamma$ via the relation $\gamma \simeq 0.63\sqrt{\Gamma_N\Delta_0}$, valid in the unitarity limit (which yields the largest $\gamma$ for a given value of $\Gamma_N$).\cite{Hirschfeld93}  
These estimates are listed in Table~\ref{table:Tlgamma}, based on the $\Delta_0$ values of Table~\ref{table:TlkappaoT}.
One can see that  $\gamma / k_B T > 13 $ for $T < 1$ K and $\gamma / \Delta_0 \simeq 0.1$ in all cases, more precisely ranging from 0.07 to 0.16.
When $\gamma$ is not negligible, there is a known correction to Eq.~\ref{eq:koT} (and Eq.~\ref{eq:koTvsDelta}), which for $\gamma / \Delta_0 \simeq 0.1$ is about 20\%,\cite{Sun95B} i.e. the clean-limit gap value is 20\% larger than estimated using Eq.~\ref{eq:koTvsDelta}. 
An error of this magnitude has little impact on the conclusions of this article.  

\begin{table}
\caption{Estimates of the impurity bandwidth, $\gamma$, from the normal state scattering rate $\Gamma_N$ = 0.52 meV, in the unitary limit.}
\label{table:Tlgamma}
\begin{tabular}{|c|c|c|c|c|}
\hline 
~$T_c (K)$~ & ~$p$~ & ~$\gamma$~(meV)~ & ~$\Delta_0$~(meV)~ & ~$\gamma/\Delta_0$~ \\
\hline
84 & ~0.188~ & 2.6 & 40 & 0.065 \\
76 & 0.203 & 2.0 & 22 & 0.088 \\
72 & 0.209 & 2.0 & 23 & 0.085 \\
68 & 0.214 & 2.2 &  28 &  0.078\\
27 & 0.252 & 1.3 & 9.5 & 0.134 \\
26 & 0.253 & 1.1 & 6.7 & 0.16 \\
\hline
\end{tabular}
\end{table}

\subsection{From the magnetic field dependence of $\kappa_0/T$}

As a crosscheck, we can also make use of the magnetic field dependence of the thermal conductivity in the $T \rightarrow 0$ limit to estimate $\gamma$ in our samples.  Unlike the zero-field thermal conductivity, which is independent of $\gamma$, the field dependence of the thermal conductivity is dependent upon the quasiparticle scattering rate.\cite{Kubert98B,Chiao99}  In a $d$-wave superconductor $\kappa_0/T$ increases with an applied magnetic field due to the Volovik effect\cite{Volovik93} and in optimally doped cuprates \cite{Chiao99,Aubin99,Hawthorn03} the thermal conductivity is well described by semi-classical calculations of the field dependence of $\kappa_0/T$ based on the Volovik effect.\cite{Kubert98B,Vekhter99}
 
\begin{figure}[th]
\centering
\resizebox{3in}{!}{\includegraphics{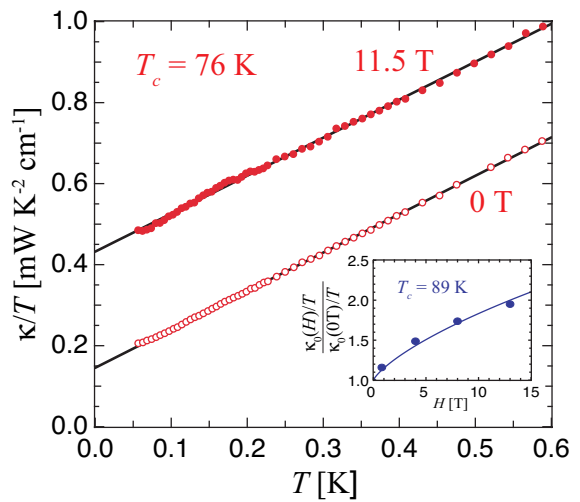}}
\caption{(colour online) Thermal conductivity in an applied field $(H // c)$ for the Tl-2201 sample with  $T_c$ = 76 K, plotted as $\kappa/T $ vs $T$.  $\kappa_0/T$ increases by a factor of $\sim 3$ in an applied field of 11.5 T. The lines are fits to Eq.~\ref{eq:kappafitTl}.  
Inset: The in-field electronic thermal conductivity $\kappa_0(H)/T$ normalized to the zero-field value  $\kappa_0(0)/T$ as a function of magnetic field in a Tl-2201 sample with $T_c = 89$ K (not used in the main study).  
The solid line is a fit to the semi-classical model (Eq.~\ref{eq:KubertkappaH}), which yields $\gamma/\Delta_0 \simeq 0.1$ (see text).}
\label{fig:kappaH}
\end{figure}

An analytic model of $\kappa_0(H)/T$ for $H \ll H_{c2}$ and ${\bf J} \perp {\bf H}$ by K{\"u}bert and Hirschfeld\cite{Kubert98B} gives:
\begin{equation}
\frac{\kappa(H)/T}{\kappa_0/T} = \frac{\rho^2}{\rho\sqrt{1+\rho^2}-\sinh^{-1}\rho},
\label{eq:KubertkappaH}
\end{equation}
where $\kappa_0/T$ is the universal value in zero field (Eq.~\ref{eq:koT}) and (for $\gamma > E_H$) 
\begin{equation}
\rho = \sqrt{6/\pi}\gamma /E_H.
\end{equation}

Here $E_H = a\hbar \sqrt{2/\pi}v_F\sqrt{H/\Phi_0}$ is the average energy shift experienced by a quasiparticle due to the Volovik effect, $a$ is a constant dependent upon the geometry of the vortex lattice (= 1/2 for a square lattice\cite{Hussey02}) and $\Phi_0$ is the quantum of flux.    Note that aside from $\gamma$, Eq.~\ref{eq:KubertkappaH} is only dependent upon known constants.  As such, we can estimate $\gamma$ from a single parameter fit to the field dependence of $\kappa_0/T$ at $H \ll H_{c2}$.

In Figure~\ref{fig:kappaH}, we show $\kappa/T$ vs $T$ for the Tl-2201 sample with $T_c = 76$ K, in $H=0$~T and 11.5 T.  We see that $\kappa_0/T$ is enhanced by a factor 3 or so. Similar field-induced enhancements of $\kappa_0/T$ are observed in all samples, by a factor which varies from 2 to 4.\cite{Hawthorn05}
In the inset, we show $\kappa_0(H)/T$ (normalized to $\kappa_0(0)/T$) vs $H$ in a separate Tl-2201 sample ($T_c = 89$ K).\cite{endnoteMac15}  
Eq.~\ref{eq:KubertkappaH} provides a reasonable fit (solid line) to the data, giving $\rho \sqrt{H}= 3.36 $T$^{1/2}$, which results in $\gamma = 3.8 $ meV, so that $\gamma / \Delta_0 \simeq 0.1$.  
This is similar in magnitude to our above estimate from normal state transport, providing a non-trivial validation for both methods of estimating $\gamma$.

We conclude that within possible corrections on the order of 20\% or so, it is legitimate to use Eqs.~\ref{eq:koT} and \ref{eq:koTvsDelta} to extract an estimate of $\Delta_0$ from a measurement of $\kappa_0/T$ in the Tl-2201 samples considered here.

\begin{acknowledgments}
We thank K. Behnia, J. Carbotte, A. Damascelli, A. Millis, M. Smith, A.-M. Tremblay and J. Wei for stimulating discussions.  This work was supported by the Canadian Institute for Advanced Research and funded by a Canada Research Chair, the Walter Sumner Foundation, the Canadian Foundation for Innovation and the National Science and Engineering Research Council of Canada.
\end{acknowledgments}

\bibliography{Tldopingbib}

\begin{thebibliography}{48}
\bibitem[{end({\natexlab{a}})}]{endnotegamma}
\bibinfo{note}{Assuming unitary impurity scattering, $\gamma/\Delta_0 \propto
  \sqrt{1/\Delta_0}$.\cite{Hirschfeld93} Thus, as $\Delta_0$ is driven to zero
  with increasing doping, the samples are by necessity driven into the limit
  $\gamma/\Delta_0 \sim 1$ as $\Delta_0 \rightarrow 0$.}
\bibitem[{end({\natexlab{b}})}]{endnoteDelta}
\bibinfo{note}{It should be cautioned that there is considerable scatter in the
  collected gap values available in literature from various techniques and
  groups, not shown here for the sake of clarity. This scatter is likely due to
  sample quality issues and variance in the modeling procedures used to
  determine $\Delta_0$ and doping, $p$. For a representative compilation of
  $\Delta_0$ see a review of the pseudogap by Timusk and Statt\cite{Timusk99}}.
\bibitem[{end({\natexlab{c}})}]{endnoteMac15}
\bibinfo{note}{This sample, while suitable for relative measurements of
  $\kappa_0(H)/T$, has an irregular geometry that does not allow for an
  accurate determination of the absolute value of $\kappa_0/T$.}
\end{thebibliography}

\end{document}